\documentclass[aps,a4paper,superscriptaddress,twocolumn,showpacs]{revtex4} 
\usepackage{graphics,epsfig}
\usepackage{setspace}
\newcommand{\be}{\begin{equation}}
\newcommand{\ee}{\end{equation}}
\newcommand{\bea}{\begin{eqnarray}}
\newcommand{\eea}{\end{eqnarray}}
\newcommand{\ba}{\begin{eqnarray*}}
\newcommand{\ea}{\end{eqnarray*}}
\newcommand{\pdag}{{\phantom{\dagger}}}

\begin{document}
\title{One-Dimensional Multi-Band Correlated Conductors and Anderson Impurity Physics}

\author{Michele Fabrizio}
\affiliation{International School for Advanced Studies (SISSA), Via Beirut 2-4, 
I-34014 Trieste,Italy}
\affiliation{International Centre for Theoretical Physics
(ICTP), P.O.Box 586, I-34014 Trieste, Italy}
\author{Erio Tosatti}
\affiliation{International School for Advanced Studies (SISSA), Via Beirut 2-4, 
I-34014 Trieste,Italy}
\affiliation{International Centre for Theoretical Physics
(ICTP), P.O.Box 586, I-34014 Trieste, Italy}

\date{\today}
\begin{abstract}

A single Anderson impurity model recently predicted, through its unstable 
fixed point, the phase diagram of a two band model correlated conductor, 
well confirmed by Dynamical Mean Field Theory in infinite dimensions. We study here 
the one dimensional version of the same model and extract 
its phase diagram in this opposite limit of reduced dimensionality. As 
expected for one dimension, the Mott metal-insulator transition at half 
filling is replaced by a dimerized insulator-undimerized Mott insulator 
transition, while away from half filling the strongly correlated 
superconductivity for inverted Hund's rule exchange in infinite dimensions is 
replaced by dominant pairing fluctuations. Many other aspects of the  
one dimensional system, in particular the field theories and their symmetries are remarkably 
the same as those of the Anderson impurity, whose importance appears enhanced.

\end{abstract}
\pacs{71.30+h, 71.10.Pm, 71.10.Fd}
\maketitle

Very substantial progress in our understanding of the Mott metal-insulator 
transition (MIT) have been made thanks to the so-called Dynamical Mean 
Field Theory (DMFT)\cite{DMFT}, a quantum analogue of the classical 
mean-field theory which treats time correlations and is exact in infinite 
dimensions ($\infty$-D). In DMFT, the approach to the MIT from the metal 
phase is accompanied by a net separation of energy scales between well 
pre-formed high energy Hubbard bands -- images of the excitations in the 
nearby Mott insulator -- and the lingering low-energy itinerant quasiparticles. 
This separation is in fact already contained in the isolated Anderson 
impurity model (AIM), where most of the spectral weight is concentrated 
in the high-energy sub-bands and only a small fraction -- describing 
quasiparticles promoted into the conduction screening-bath -- remains 
close to the chemical potential. This is of course unsurprising since 
in $\infty$-D DMFT maps the lattice model of interacting electrons onto 
an AIM supplemented by a self-consistency condition\cite{DMFT}. Irrespective of 
whether this mapping is merely a trick to solve the lattice model in 
$\infty$-D or whether it hides perhaps a more fundamental aspect of the physics close to a 
MIT, this does suggest that some of the strongly correlated lattice properties 
could be directly inferred by the AIM itself, even without self-consistency. 
This route was recently explored to anticipate the anomalous properties 
near the MIT of a two-band Hubbard conductor on the basis of the phase 
diagram and in particular of the unstable fixed point of a two-orbital 
AIM\cite{Deleo}. All the predicted properties, including strongly correlated 
superconductivity near the Mott transition\cite{Capone02} were later 
confirmed by full DMFT\cite{Capone04}. Despite that success, it would still 
seem hazardous to suggest that the properties of a single AIM have generally 
anything to do with the actual behavior of the model lattice conductor 
away from $\infty$-D, least of all in the opposite extreme of one dimension (1D).
We show in this Letter that, apart from some obvious differences related 
to dimensionality, the phase diagram of the model does not change significantly 
in 1D, where therefore the AIM physics appears to remain significant. 

We consider the two-band Hamiltonian near half-filling and in 1D
\bea
\hat{H} &=& -t\,\sum_{a=1}^2 \, \sum_{i\,\sigma} \left(
c^\dagger_{i,a\sigma}c^\pdag_{i+1,a\sigma} + H.c. \right)
+ \frac{U}{2}\,\sum_i\, \left(\hat{n}_i-2\right)^2 \nonumber \\
&& - \frac{J}{4}\,\sum_{a\not = b}\, \sum_{i\,\sigma\,\sigma'}  
c^\dagger_{i,a\sigma}c^\dagger_{i,b\sigma'}c^\pdag_{i,a\sigma'}
c^\pdag_{i,b\sigma},\label{Ham}
\eea
where $c^\dagger_{i,a\sigma}$($c^\pdag_{i,a\sigma}$) 
creates(annihilates) an electron at site $i$, 
in orbital $a=1,2$ with spin $\sigma=\uparrow,\downarrow$, 
and $\hat{n}_i=\sum_{a\sigma} c^\dagger_{i,a\sigma}c^\pdag_{i,a\sigma}$  
is the electron density at site $i$, and $U \gg |J|$ is an on-site
Coulomb repulsion. This Hamiltonian has recently been discussed as relevant 
to the novel doped metal-phthalocyanine conductors.\cite{Tosatti} 
For $J=0$ Eq.~(\ref{Ham}) describes an SU(4) Hubbard model, 
analysed {\em e.g.} in Ref.~\cite{Azaria}. 
A finite value of $J$ lowers the symmetry down to  
U(1)$\times$SU(2)$\times\left({\rm U(1)}\times{\rm Z}_2\right)$, 
where U(1) refers to charge, SU(2) to spin and 
$\left({\rm U(1)}\times{\rm Z}_2\right)$ to the flavour (orbital) sector. 
We stress here that the single AIM shares identically this same 
symmetry, a point which we will return to further down. 
Two electrons on the same site can form either a spin triplet, 
with energy $J/4$, an inter-orbital singlet, with energy $-3J/4$, or 
two intra-orbital singlets with energy $-J/4$. Therefore $J<0$ favors 
the spin triplet while $J>0$ the inter-orbital singlet.
Actually $J>0$, (``inverted Hund's rule exchange''), provides 
a pairing mechanism in the Cooper channel $c^\dagger_{i,1\uparrow}c^\dagger_{i,2\downarrow} + 
c^\dagger_{i,2\uparrow}c^\dagger_{i,1\downarrow}$. Pairing is impeded by 
the repulsion $U$, so that the {\sl bare} scattering amplitude in the 
inter-chain singlet channel, $A=U-J/2$, is attractive only in the unrealistic case 
of $J>2U>0$, apparently excluding superconductivity despite the pairing
mechanism provided by $J>0$. As shown in DMFT \cite{Capone04}, this n\"{a}ive expection 
is actually wrong, at least in $\infty$-D. 
A superconducting pocket appears near the half-filled Mott insulator, 
$U\sim t \gg J$,  moreover with a hugely enhanced superconducting gap 
with respect to the $U=0$ BCS gap value. This surprising result had in fact 
been foreshadowed by the single AIM study\cite{Deleo} 
whose phase diagram displays an unstable fixed point at $0<J_*\simeq T_K$, 
where $T_K$ is the Kondo temperature, which separates a Kondo screened phase 
for $J<J_*$ from an unscreened phase for $J>J_*$.
Translated into DMFT, the AIM Kondo temperature becomes the 
quasiparticle coherent bandwidth, vanishing at the MIT. 
This implies that the AIM onto which the metallic lattice model 
maps in $\infty$-D must necessarily cross the unstable fixed point 
$J\sim T_K$ before the MIT. The speculation\cite{Deleo} 
that the lattice model would respond to the local instability by spontaneously developing a 
bulk order parameter in the inter-orbital singlet Cooper channel
was fully confirmed by DMFT\cite{Capone04}.   
  
We turn now to study the same model in 1D. As usual it is convenient here 
to represent (\ref{Ham}) within bosonization.\cite{GN&T} The Fermi fields 
around the right (R), $+k_F$, and left (L), $-k_F$, 
Fermi points are expressed as $\psi_{R(L),a\sigma}(x)\sim \exp\left[\pm ik_F\,x -i\sqrt{\pi}
\left(\Theta_{a\sigma}(x)\mp \Phi_{a\sigma}(x)\right)\right]$, where $\Phi_{a\sigma}(x)$ 
and $\partial_x \Theta_{a\sigma}(x)$ are conjugate Bose fields. We introduce 
the linear combinations $\Phi_c = \left(
\Phi_{1\uparrow}+\Phi_{1\downarrow} +\Phi_{2\uparrow}+\Phi_{2\downarrow}\right)/2$, 
$\Phi_s = \left(
\Phi_{1\uparrow}-\Phi_{1\downarrow} +\Phi_{2\uparrow}-\Phi_{2\downarrow}\right)/2$, 
$\Phi_f = \left(
\Phi_{1\uparrow}+\Phi_{1\downarrow} -\Phi_{2\uparrow}-\Phi_{2\downarrow}\right)/2$, and 
$\Phi_{sf} = \left(
\Phi_{1\uparrow}-\Phi_{1\downarrow} -\Phi_{2\uparrow}+\Phi_{2\downarrow}\right)/2$, which 
describe respectively the total charge, the total spin, the relative charge 
and the relative spin density fluctuations. In this representation the interaction involves 
only bilinears of $\cos\sqrt{4\pi} \Phi_{n}$ and $\cos\sqrt{4\pi} \Theta_n$, $n=c,s,f,sf$, 
which in turn can be expressed as 
\bea
\cos\sqrt{4\pi} \Phi_n &=& -i\frac{\pi\alpha}{2}\left(
\xi_{R,n}\xi_{L,n} + \zeta_{R,n}\zeta_{L,n}\right),\label{cosphi}\\
\cos\sqrt{4\pi} \Theta_n &=& i\frac{\pi\alpha}{2}\left(
\xi_{R,n}\xi_{L,n} - \zeta_{R,n}\zeta_{L,n}\right),\label{costheta}
\eea
where $\xi_{R(L),n}$ and $\zeta_{R(L),n}$ are Majorana fermions and $\alpha$ is a 
cutoff distance. These fermions can be used to introduce 
eight two-dimensional classical Ising models, each one 
in principle characterized by a {\sl mass} $m\sim (T-T_c)$, $m<0$ and $m>0$ meaning ordered 
and disordered phases, and $m=0$ the critical point.  
In this way the U(1) charge sector is represented by two identical Ising models with 
mass $m_c$,
the doublet $(\xi_{R,c},\xi_{L,c})$ and $(\zeta_{R,c},\zeta_{L,c})$;  
the SU(2) spin sector by three identical Ising models (mass $m_s$),  
the triplet  $(\xi_{R,s},\xi_{L,s})$, $(\zeta_{R,s},\zeta_{L,s})$ and    
$(\xi_{R,sf},\xi_{L,sf})$; the U(1) flavour sector by two 
identical Ising copies (mass $m_f$), the doublet $(\xi_{R,f},\xi_{L,f})$ and  
$(\zeta_{R,f},\zeta_{L,f})$; and finally the remaining flavour Z$_2$ by 
a single Ising model (mass $m_0$), the singlet $(\zeta_{R,sf},\zeta_{L,sf})$.\cite{notac}  
Without interaction all Ising models are critical. The interaction 
induces marginally relevant couplings between them which  
might spontaneously generate finite masses. Indeed, by a fermion two-loop 
renormalization group (RG) analysis, we find that the Hamiltonian 
generally flows to strong coupling fixed points which allow a simple mean-field 
description in terms of finite average values of $\langle \xi_{R,n}\xi_{L,n}\rangle$ 
and $\langle \zeta_{R,n}\zeta_{L,n}\rangle$, $n=c,s,f,sf$, which preserve 
all continuous symmetries. Within this same description the phase diagram of (\ref{Ham}) 
turns out to be characterized by simply identifying the relative signs of the 
masses $m_i$, $i=c,s,f,0$, while the overall sign has a physical 
meaning only in the spontaneoulsy dimerized phase, see below.\cite{Balents}

\begin{figure}[t] 
\includegraphics[width=8cm]{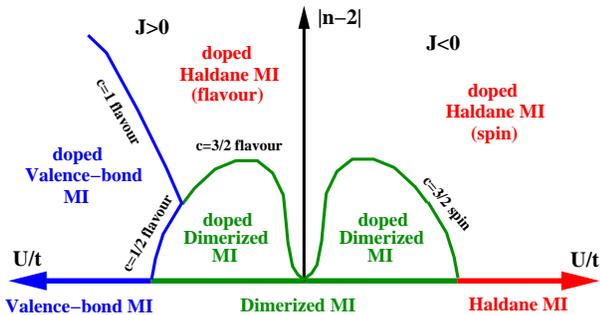} 
\caption{Phase diagram of model (\ref{Ham}) near half-filling as a function 
of doping $|n-2|$ and of $U/t$ at fixed $|J|/U\ll 1$ for 
$J<0$ (right panel) and $J>0$ (left panel).
At half-filling the system is always insulating and displays by increasing $U/t$ 
an Ising transition ($c=1/2$ flavour) from a Dimerized Mott insulator (MI) 
to a Valence-bond MI for $J>0$; and a $c=3/2$ transition to a S=1 Haldane MI 
for $J<0$ where the spin is gapless ($c=3/2$ spin). All the insulating phases evolve 
upon doping into metal with gaps in all noncharge sectors. For the doped Haldane MI 
the additional label (flavour) or (spin) indicates that the phase can be viewed as 
the natural evolution away from half-filling of a Haldane chain built of either 
spin or flavour triplets. The transition lines bewteen each different metallic phase 
are identified by the central charge $c$ and by the sector involved, 
spin or flavour. \label{phd} } 
\end{figure}

We start from half filling, $\langle \hat{n}_i\rangle = 2$, and analyse the 
phase diagram for increasing $U/t$ keeping for simplicity a fixed ratio $U/|J|\gg 1$. 
At weak coupling, $U/t \ll 1 $, the Hamiltonian flows under RG to an SO(8) Gross-Neveu model, 
which describes a spontaneously dimerized insulator with
gaps in the whole excitation spectrum. It is 
known that the SU(4) Hubbard model, {\sl i.e.} (\ref{Ham}) with $J=0$, 
dimerizes at half-filling\cite{Azaria,Boulat}, and that a 
small $|J|\ll U$ cannot destabilize this gapped phase.\cite{Azaria2} 
This phase is characterized by all masses having the same sign, 
with the overall sign reflecting the broken translational symmetry. 
Eventually, though,
this dimerized phase cannot survive indefinitely for large 
$U/t$. When $U\gg t$, two electrons localize at each site in a configuration 
optimizing the on-site exchange $J$. In particular for $J<0$, conventional Hund's rules,  
the model effectively reduces to a spin S=1 Heisenberg chain, 
still gapped everywhere in the spectrum but not dimerized\cite{Haldane}.   
For inverted Hund's rules, $J>0$ , the singlet configuration 
\be
\sqrt{\frac{1}{2}}\, 
\left[c^\dagger_{i,1\uparrow}c^\dagger_{i,2\downarrow} + 
c^\dagger_{i,2\uparrow}c^\dagger_{i,1\downarrow}\right]\,|0\rangle,
\label{singlet}
\ee
is favored and the ground state is akin to a collection of local singlets, 
a kind of local valence-bond Mott insulator, still gapped but not dimerized. 
We conclude that upon increasing $U/t$ the dimerization must disappear for 
either sign of $J$. In fact, whereas at weak coupling at for $J=0$ the 
dimerization-induced gaps in the spin and flavour sectors follow the 
BCS-like behavior of the charge gap, the latter continues to increase 
monotonically as $U$ increases while the former gaps reach a maximum, 
approximately when $U\sim 5t$, and then start dropping as $t^2/U$ for 
$U\gg t$\cite{Azaria,Boulat}. This decoupling of charge from spin and 
orbital modes is a 1D remnant of the MIT and seems quite sharp\cite{Azaria}. 
The weak exchange $|J|/U\ll 1$ , irrelevant in the weak coupling dimerized 
phase, eventually turns in strong coupling to a relevant perturbation 
able to suppress dimerization. Thus one might expect that this could 
occur only for very large $U$, when the spin gap induced by dimerization 
becomes small of order $|J|$. However it cannot be excluded that the demise 
of dimerization could even take place for smaller $U$, say for $U<5t$.
The two-loop RG equations moreover suggest for $J<0$ a $c=3/2$ spin-SU(2)$_2$ 
critical point where the triplet mass $m_s$ changes sign\cite{notab}, signaling a transition 
from the dimerized insulator to the Haldane spin-1 chain Mott insulator.
For $J>0$ the transition is instead predicted to occur through a $c=1/2$ Ising critical
point where the singlet mass $m_0$ crosses zero, signaling the transition from a dimerized 
to the valence-bond Mott insulator, see Fig.~\ref{phd}. 

Let us introduce doping, moving away from half filling. Dropping the 
Umklapp terms from the weak coupling RG equations, the interaction flows, 
for either sign of $J$, towards a fixed point 
\be
\hat{H}_{int} \rightarrow - g_*\, \int dx\, \Delta^\dagger(x)\,\Delta(x),
\label{fixed-point}
\ee
where 
\be
\Delta^\dagger = 
\psi^\dagger_{R,1\uparrow}\psi^\dagger_{L,2\downarrow}   +
\psi^\dagger_{L,2\uparrow}\psi^\dagger_{R,1\downarrow} -  
\psi^\dagger_{L,1\uparrow}\psi^\dagger_{R,2\downarrow} -    
\psi^\dagger_{R,2\uparrow}\psi^\dagger_{L,1\downarrow}
\label{Cooper-H}
\ee
which is a spin-singlet, flavour-singlet but space-odd pairing operator\cite{Azaria2,nota}. 
This fixed point interaction has a dynamically enlarged 
SU(4) symmetry\cite{Azaria2,Balents}, unlike the original model (\ref{Ham}), and 
realizes a doped Haldane chain, the two 1/2-spin constituents forming singlet bonds,  
one to its right and the other to its left.\cite{Affleck} Moving from $J<0$ to $J>0$ 
interchanges the spin sector and the flavour sector, which includes the doublet and 
the singlet forming a degenerate triplet, again a dynamically enlarged SU(2)$_2$ 
flavour symmetry. The Ising masses satisfy  $m_c=0$, $m_f\,m_0>0$ but $m_s\, m_f <0$.  
The pairing correlation function 
$\langle \Delta(x) \Delta^\dagger(0)  \rangle \sim (1/x)^{1/2K_c}$ where $K_c$ 
is the Luttinger liquid exponent of the gapless U(1) charge sector. There are also 
power-law decaying $4k_F$ correlation functions with exponent $2K_c$ 
which involve the density-wave operators  
$\exp\left(\pm i 4\,k_F\,x \pm i\sqrt{4\pi}\Phi_c\right)\, \cos\sqrt{4\pi}\Phi_{s(f)}$   
and $\exp\left(\pm i 4\,k_F\,x \pm i\sqrt{4\pi}\Phi_c\right)\, \cos\sqrt{4\pi}\Theta_{sf}$.    
If $K_c>1/2$ the superconducting fluctuations dominate over the 
$4k_F$ density-wave ones, and the opposite for $K_c<1/2$. Since 
the model has an insulating phase at quarter filling\cite{Azaria3}, by 
standard arguments\cite{Schulz} we expect $K_c\geq 1/4$. Hence the 
pairing susceptibility in channel (\ref{Cooper-H}) always diverges faster than 
for free fermions. 

Revealing as it is, this weak coupling analysis 
is not fully satisfying as it implies an unphysically abrupt change of sign 
of $m_s$ at the slightest density deviation from half-filling. 
A better approach near half filling may be a two-cutoff RG scheme, namely 
running at first the RG as if for half filling 
until reaching an energy scale of the order of the chemical potential shift, 
and only at this point dropping the Umklapp terms. Doing this we find that the doped
dimerized Mott insulator turns to a metallic phase, $m_c=0$, all other masses 
retaining their sign. Here the dimer order parameter is zero 
but its correlation function decays slowly with a power-law exponent $K_c/2$. 
This agrees with a similar analysis by Boulat\cite{Boulat} and 
leads us to propose the phase diagram of Fig.~\ref{phd} where the doped dimerized 
insulator transform into the doped Haldane phase by a $c=3/2$ critical line\cite{notab} which 
involves the spin or the flavour sector for $J<0$ and $J>0$, respectively. 
In addition we expect, by similar arguments, that the valence bond Mott insulator at 
large $U/t$ with $J>0$ transforms upon doping to a metal phase, with $m_c=0$, 
$m_s\, m_f >0$  but $m_0\, m_f <0$. This phase is identified by a fixed point 
interaction of exactly the same form as (\ref{fixed-point}) 
with the pairing operator corresponding 
to the singlet configuration (\ref{singlet}), namely  
\be
\Delta^\dagger = 
\psi^\dagger_{R,1\uparrow}\psi^\dagger_{L,2\downarrow}   +
\psi^\dagger_{L,2\uparrow}\psi^\dagger_{R,1\downarrow} +  
\psi^\dagger_{L,1\uparrow}\psi^\dagger_{R,2\downarrow} +    
\psi^\dagger_{R,2\uparrow}\psi^\dagger_{L,1\downarrow},
\label{Cooper-VB}
\ee
which is still a spin-singlet but it is now no longer invariant under the flavour SU(2). 
The pairing correlation function in this channel still 
decays with exponent $1/2K_c$, and again there are competing 
$4k_F$ density wave fluctuations with exponent $2K_c$. Therefore we argue that 
the Ising critical point at half-filling for $J>0$ and large $U/t$ extends to a 
critical $c=1/2$ line away from half-filling, as in Fig.~\ref{phd}. This 
line should merge into the $c=3/2$ flavor SU(2)$_2$ critical line, out of which a $c=1$ 
U(1) critical line must emerge. Along this line the flavor doublet mass, $m_f$, 
changes sign. This scenario is compatible with our expectation far away from half-filling. 
Indeed, if we keep the ratio $J/U>0$ fixed and increase $U/t$ we arrive at a situation 
where $J\gg t$. Here, whenever two electrons occupy the same site, they are forced 
into the singlet configuration (\ref{singlet}). This constraint can be implemented 
by a projector leading to a model which was numerically 
analysed in Ref.~\cite{Parola} close to quarter filling. The numerical results are 
compatible with the existence of the doped valence-bond phase. Since the weak 
coupling phase is instead the doped Haldane chain, 
we conclude that there is a transition by increasing $U/t$ at fixed $J/U>0$ 
whose criticality belongs to the U(1) universality class. 
Out of all these arguments we finally draw the qualititative phase diagram of Fig.~\ref{phd}. 

We note the presence throughout the phase diagram (except of course at 
half filling) of a singular spin-singlet Cooper pairing susceptibility either 
in channel (\ref{Cooper-H}), within the doped Haldane insulator, or in channel 
(\ref{Cooper-VB}), within the doped valence-bond insulator. The latter, as discussed 
at the beginning, is most unexpected, emerging only for sufficiently strong repulsion $U$.  
We note that, if $0<J\ll t$ were fixed, we still would expect at $U=0$  
singular pairing susceptibilities below an exponentially small energy scale of the 
order of the spin and flavour gaps. Remarkably, we find that 
the role of a large $U$ is to raise this energy scale by increasing the 
magnitude of the gaps, which in turns implies that $U$ effectively enhances 
pairing fluctuations. 

Finally we consider the competition between Cooper pairing and  
$4k_F$ density-wave fluctuations in the doped valence bond and in the Haldane phases as 
we approach the Mott insulator at half filling. The insulating behavior is 
driven at weak coupling by $4k_F$-Umklapp terms of the form 
$\cos\sqrt{4\pi}\Phi_c\, \cos\sqrt{4\pi}\Phi_s\left(\cos\sqrt{4\pi}\Phi_f\right)
\left(\cos\sqrt{4\pi}\Theta_{sf}\right)$. Since the spin and flavour sectors 
remain gapped away from half filling, one is tempted to conclude that 
$K_c\to 1$  on approaching half filling\cite{Schulz}. 
However, both Haldane and valence bond 
Mott insulators require a large $U/t$, and the above weak-coupling argument 
might not be correct. One may argue that the decoupling of charge from 
the other sectors as $U/t$ increases may be modelled by adding an $8k_F$-Umklapp 
$\cos\sqrt{16\pi}\Phi_c$\cite{Azaria}. If alone this term would rather suggest a MIT  
at half filling with $K_c\to 1/4$. 
However it was found numerically that in the half-filled SU(4) Hubbard model  
the charge-2 Majorana fermions, 
$\xi_{R(L),c}$ and $\zeta_{R(L),c}$, remain coherent excitations even at large $U$, 
even though their 
energy is only slightly below the two-particle continuum, merging into the latter only for 
$U/t\to \infty$. This suggests that $K_c$ tends always to 1  
as the density approaches half filling, even though a larger $U$ implies a sharper
crossover from $K_c\simeq 1/4$ to $K_c=1$. Hence we conclude that sufficiently close to 
the MIT the Cooper channel pairing dominates over $4k_F$ fluctuations.    

We can finally return to our original motivation and discuss differences/analogies 
with the phase diagram of the same model suggested by the AIM and verified by DMFT 
in infinite dimensions for $J>0$.  At half filling we always find in 1D an insulator
because of perfect nesting, generally absent in higher dimensions.
Hence the MIT of $\infty$-D is replaced in 1D by an Ising transition between two 
insulators, the band-like dimerized insulator (driven by nesting) 
and the strong-coupling undimerized valence-bond Mott insulator. Upon doping, we still expect 
by increasing $U/t$ an Ising transition from a doped dimerized insulator into a 
doped valence-bond insulator, the latter characterized by a singular pairing 
susceptibility. This is the exact analog of what was found in infinite dimensions. 
Even more surprising is the role of the singlet Ising sector $(\zeta_{R,sf},\zeta_{L,sf})$ 
which becomes critical at the transition. As previously mentioned, the 
behavior of the lattice model in $\infty$-D is controlled near the MIT by the 
unstable fixed point of the AIM onto which the lattice model maps. 
In turn this unstable fixed point can be interpreted as the free boundary condition 
fixed point separating the two fixed boundary conditions just in the same Ising sector, 
and that is the natural generalization to a boundary problem of the Ising critical point 
which we uncovered in 1D. It is now clear that both the 1D and the single 
impurity analysis make use of the same field theoretical scheme, the 
non-abelian bosonization, where 
the Ising sector emerges naturally from embedding the 
flavour SU(2)$_2$ into U(1). 
Suggestively, this appears to be the reason why the behavior in 
one dimension and in infinite dimensions are so similar. At this stage we  
cannot say whether this similarity is a mere accident or not. It certainly does
encourage the speculation that the physics of a single AIM may play a more 
fundamental role in the description of strongly correlated metals in any dimensions, 
at least at intermediate energy/temperature scales before full bulk coherence settles in. 

We are grateful to A.A. Nersesyan and P. Lecheminant for useful advices. This 
work has been partly supported by MIUR COFIN 2003 and by INFM/FIRB RBAU017S8R.


\begin{thebibliography}{30}
\bibitem{notac} For the U(1) sectors it is more appropriate to speak of 
Askin-Teller models.
\bibitem{notab} We cannot exclude that the transition could in reality be first order, 
even though our fermion RG does include all marginal interactions.
\bibitem{nota} In Ref.~\cite{Azaria2} this Cooper operator was misleadingly 
identified as a spin-triplet.
\end{thebibliography}
\end{document}